\documentclass[aps,prl,twocolumn,superscriptaddress,longbibliography]{revtex4-2}

\usepackage{graphicx}
\usepackage{amsmath}
\usepackage{upgreek}
\usepackage{setspace}
\usepackage{hyperref}
\hypersetup{colorlinks,breaklinks,
            urlcolor=[rgb]{0,0,0.64},
            linkcolor=[rgb]{0,0,0.64},
            citecolor=[rgb]{0,0,0.64},
            filecolor=[rgb]{0,0,0.64}}
\usepackage{nccmath}
\usepackage{amssymb}

\newcommand{\MIT}{Massachusetts Institute of Technology, Department of Physics, Cambridge, MA 02139, USA.}
\newcommand{\Cal}{University of California at Berkeley, Department of Chemistry, Berkeley, CA 94720, USA.}
\newcommand{\ICQM}{International Center for Quantum Materials, School of Physics, Peking University, Beijing 100871, China.}
\newcommand{\SLML}{Songshan Lake Materials Laboratory, Dongguan, Guangdong 523808, China.}
\newcommand{\SSRL}{Stanford Synchrotron Radiation Lightsource, SLAC National Accelerator Laboratory, Menlo Park, CA 94025, USA.}
\newcommand{\SIMES}{Stanford Institute for Materials and Energy Sciences, SLAC National Accelerator Laboratory and Stanford University, Menlo Park, CA 94025, USA.}
\newcommand{\StanfordAP}{Department of Applied Physics, Stanford University, Stanford, CA 94305, USA.}
\newcommand{\SSRF}{Shanghai Synchrotron Radiation Facility, Shanghai Advanced Research Institute, Chinese Academy of Sciences, Shanghai 201204, China.}
\newcommand{\APS}{Advanced Photon Source, Argonne National Laboratory, Lemont, IL 60439, USA.}
\newcommand{\SLAC}{SLAC National Accelerator Laboratory, Menlo Park, CA, USA.}
\newcommand{\MIPT}{Moscow Institute of Physics and Technology, Institutsky per. 9, Dolgoprudny, 141701 Russia.}
\newcommand{\Leipzig}{Institute for Theoretical Physics,
University of Leipzig, 
Brüderstrasse 16,
04103 Leipzig, Germany.}
\newcommand{\ITAE}{Institute for Theoretical and Applied Electrodynamics, Russian Academy of Sciences, Moscow, 125412, Russia.}
\newcommand{\Cornell}{CHESS, Cornell University, Ithaca, NY 14853, USA.}
\newcommand{\Clemson}{Department of Physics and Astronomy, Clemson University, Clemson, SC 29631, USA.}

\newcommand{\papertitle}{Unconventional hysteretic transition in a charge density wave}

\begin{document}

\title{\papertitle}

\author{B.~Q.~Lv}
\thanks{These authors contributed equally to this work.}
\affiliation{\MIT}
\author{Alfred~Zong}
\thanks{These authors contributed equally to this work.}
\affiliation{\MIT}
\affiliation{\Cal}
\author{D.~Wu}
\affiliation{\ICQM}
\affiliation{\SLML}
\author{A.V.~Rozhkov}
\affiliation{\ITAE}
\author{Boris~V.~Fine}
\affiliation{\MIPT}
\affiliation{\Leipzig}
\author{Su-Di~Chen}
\affiliation{\StanfordAP}
\affiliation{\SIMES}
\author{Makoto~Hashimoto}
\affiliation{\SSRL}
\author{Dong-Hui Lu}
\affiliation{\SSRL}
\author{M.~Li}
\affiliation{\SSRF}
\author{Y.-B.~Huang}
\affiliation{\SSRF}
\author{Jacob P. C. Ruff}
\affiliation{\Cornell}
\author{Donald~A.~Walko}
\affiliation{\APS}
\author{Z.~H.~Chen}
\affiliation{\SSRF}
\author{Inhui~Hwang}
\affiliation{\APS}
\author{Yifan~Su}
\affiliation{\MIT}
\author{Xiaozhe~Shen}
\affiliation{\SLAC}
\author{Xirui~Wang}
\affiliation{\MIT}
\author{Fei~Han}
\affiliation{\MIT}
\author{Hoi~Chun~Po}
\affiliation{\MIT}
\author{Yao Wang}
\affiliation{\Clemson}
\author{Pablo~Jarillo-Herrero}
\affiliation{\MIT}
\author{Xijie~Wang}
\affiliation{\SLAC}
\author{Hua~Zhou}
\affiliation{\APS}
\author{Cheng-Jun~Sun}
\affiliation{\APS}
\author{Haidan~Wen}
\affiliation{\APS}
\author{Zhi-Xun~Shen}
\affiliation{\StanfordAP}
\affiliation{\SIMES}
\author{N.~L.~Wang}
\affiliation{\ICQM}
\author{Nuh~Gedik}
\email[Correspondence to: ]{gedik@mit.edu}
\affiliation{\MIT}

\begin{abstract}
Hysteresis underlies a large number of phase transitions in solids, giving rise to exotic metastable states that are otherwise inaccessible. Here, we report an unconventional hysteretic transition in a quasi-2D material, EuTe$_\text{4}$. By combining transport, photoemission, diffraction, and x-ray absorption measurements, we observed that the hysteresis loop has a temperature width of more than 400~K, setting a record among crystalline solids. The transition has an origin distinct from known mechanisms, lying entirely within the incommensurate charge-density-wave (CDW) phase of EuTe$_4$ with no change in the CDW modulation periodicity. We interpret the hysteresis as an unusual switching of the relative CDW phases in different layers, a phenomenon unique to quasi-2D compounds that is not present in either purely 2D or strongly-coupled 3D systems. Our findings challenge the established theories on metastable states in density wave systems, pushing the boundary of understanding hysteretic transitions in a broken-symmetry state.
\end{abstract}

\date{\today}

\maketitle

Hysteresis is a history-dependent response of a system when subjected to an external perturbation. This phenomenon is ubiquitous in physics, biology, and even economics. In condensed matter, it forms the pillar of modern technologies ranging from memory device \cite{Kahn1998} to functional polymers \cite{Hu2011}. In a hysteretic transition, phase change occurs at different values of a control parameter when it is swept in opposite directions. The complex interplay of states involved in these transitions provides a fertile ground for realizing novel metastable phases that do not normally exist in equilibrium \cite{Stojchevska2014,Morrison2014,Park2013}. 

A model system for observing metastable states manifested through a hysteretic transition is offered by compounds that exhibit a charge density wave (CDW). It was realized early on that CDWs possess an array of nonequilibrium states accessible by application of an electric field or variation of temperature \cite{Gruner1988}, and more recently it was shown that a ``hidden'', long-lasting state could form following photoexcitation by a femtosecond laser pulse \cite{Stojchevska2014,Gerasimenko2019}. These metastable phenomena are well understood in terms of the interaction among defects, electron condensate, and lattice superstructure, which leads to a slight deformation of the CDW periodicity or a phase slip between adjacent CDW domains \cite{ZZ2004,Ma2016,Cho2016,sergei2004}. The metastable states hold great potential in applications from nonvolatile memory to ultrafast switch \cite{Ogawa2002,Mraz2021}. They also provide experimentally accessible platforms to tackle perennial questions on jamming transitions and glass dynamics \cite{Littlewood1988,Staresinic2008,Gerasimenko2019}. Here, we report an unusual type of metastability in a layered CDW compound, EuTe$_4$, where a giant thermal hysteresis is observed to span more than 400~K, the largest value known in crystalline solids. Unlike other CDW systems, the hysteresis manifests solely in the order parameter amplitude without any effect on the modulation wavevector. Our study raises new possibilities of hysteretic behavior in the solid state, offering an important dimension for how an order parameter evolves in a symmetry-broken phase.

\begin{figure*}[htb!]
	\includegraphics[width=1\textwidth]{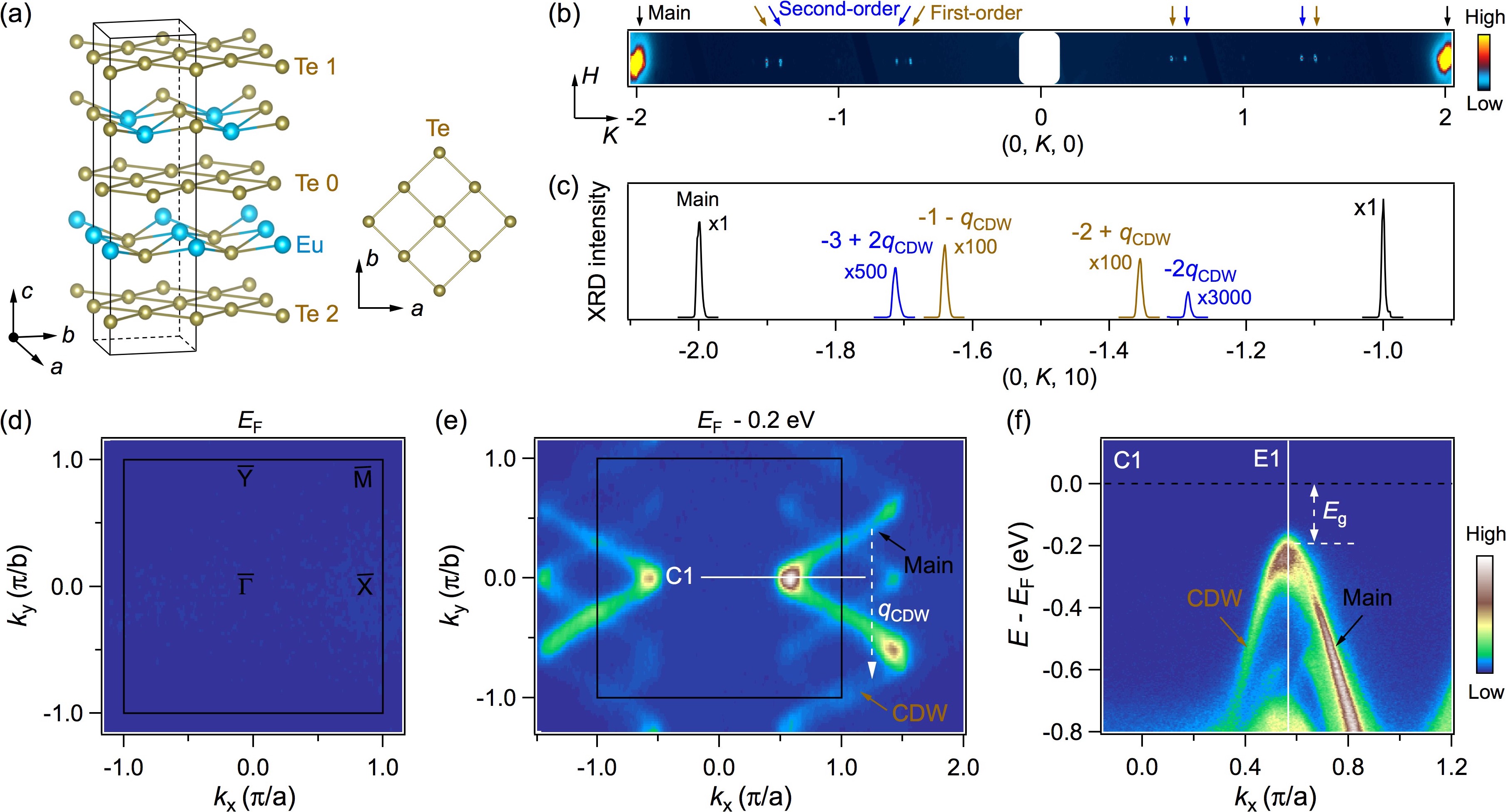}
	\caption{Incommensurate CDW in EuTe$_\text{4}$. (a)~Crystal structure of EuTe$_4$. Black lines denote the primary unit cell, which contains two motifs of the Te square lattice (\textit{right panel}): bilayer (Te~1 and Te~2) and monolayer (Te~0). (b)~Room temperature x-ray diffraction pattern taken with 30.6\,keV photon energy in transmission mode. (c)~Room temperature high-resolution x-ray intensity cut along the (0,\,$K$,\,10) direction, taken with 11\,keV photons in reflection geometry. (d)~Room temperature Fermi surface map, showing no intensity contrast. (e)~ARPES constant energy contour at 0.2\,eV below the $E_F$, measured at $T = 20$\,K. (f)~Band dispersions at 55\,K, corresponding to the C1 cut. Photon energy used was 90\,eV in (d)(e) or 24\,eV in (f).}
\label{fig:intro}
\end{figure*}

\begin{figure*}[htb!]
	\includegraphics[width=1.\textwidth]{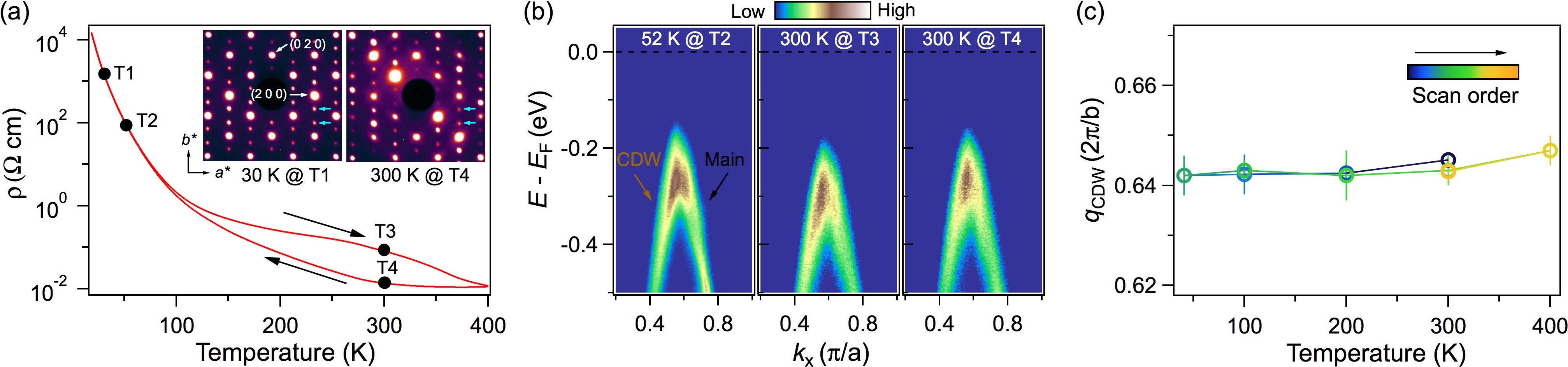}
	\caption{Temperature evolution of lattice and electronic structures. (a)~Temperature-dependent electrical resistivity with currents along the $a$-axis. Black arrows indicate the temperature sweep direction. \emph{Inset}: electron diffractions at 30\,K and 300\,K. Cyan arrows indicate the CDW superlattice peaks. (b)~Photoemission intensity with 24\,eV photons along $\bar{\Gamma}$--$\bar{\text{X}}$ at three points along the hysteresis loop [T2--T4 in (a)]. (c)~Temperature evolution of the CDW wavevector from x-ray diffraction. Error bars represent the difference in $q_\text{CDW}$ as obtained from two separate CDW peaks, $(0,-1\!-\!q_\text{CDW},10)$ and $(0,-2\!+\!q_\text{CDW},10)$.}
\label{fig:nolockin}
\end{figure*}

EuTe$_4$ possesses a layered structure where the CDW modulation originates from nearly square-shaped Te sheets [Fig.\,\ref{fig:intro}(a)] \cite{Wu2019}. We begin by characterizing the structural signatures of the CDW. The electron diffraction pattern of EuTe$_4$ in the $(HK0)$ plane reveals a unidirectional CDW along the $b$-axis. As shown in the inset of Fig.~\ref{fig:nolockin}(a), the CDW is marked by a pair of satellite peaks flanking the crystal Bragg peaks, where the CDW modulation is at $q_\text{CDW}\approx\frac{2}{3}b^*$, where $b^*\equiv \frac{2\pi}{b}$. To determine whether the CDW is commensurate to the lattice, we performed high-momentum-resolution x-ray diffraction along the $K$ cut [Fig.~\ref{fig:intro}(b)(c)]. The single CDW peak in electron diffraction is split into two distinct satellites. They can be attributed to the first and second harmonics of the CDW peaks with a modulation wavevector $q_\text{CDW}=0.643(3)b^*$. As no such splitting is allowed for a commensurate CDW at $\frac{2}{3}b^*$, we hence conclude that the density wave is not commensurate with a small integer denominator. Based on the width of the high-resolution diffraction peak [Figs.\,\ref{fig:intro}(c) and S1], we estimated the CDW correlation length to be at least 77\,nm within the Te plane and 107\,nm perpendicular to the planes, confirming the long-range nature of the superlattice modulation.

The CDW formation also leads to a spectral gap of the entire Fermi surface, as revealed by angle-resolved photoemission spectroscopy (ARPES) [Fig.~\ref{fig:intro}(d)]. The gap is not uniform along the Fermi surface contour, peaking near the $\bar{\Gamma}$--$\bar{\text{X}}$ cut. At 55\,K, the maximum gap value is $E_g=196(7)$\,meV, measured by the distance from the leading edge of the valence band to the chemical potential [Fig.~\ref{fig:intro}(f)]. This gap size yields 646\,K as a lower bound for the mean-field transition temperature \cite{Gruner1994}; indeed, the actual CDW transition occurs well above 400\,K, the highest temperature attainable in most experiments conducted. The gap opening is accompanied by the appearance of a folded CDW band, visible in the $\bar{\Gamma}$--$\bar{\text{X}}$ cut and in the constant energy map [Fig.~\ref{fig:intro}(e)(f)] (see ref.~\cite{Remark1} for a detailed description of band topology). The sharp appearance of the folded band and a complete absence of spectral weight within the energy gap suggest a well-defined CDW order with long-range phase coherence, corroborating the diffraction experiments.

While diffraction and photoemission measurements reveal a prototypical CDW in EuTe$_4$, an unusual hysteresis is seen in the electrical resistivity between 80\,K and 400\,K, a temperature range within the CDW state [Fig.\,\ref{fig:nolockin}(a)]. The hysteresis is reproduced across different sample batches and orientations [Fig.\,S6(a)]. It is also a minor loop, where the upper onset temperature of the major loop is higher than 500\,K [Fig.\,S6(b)], which is the maximum temperature we could access experimentally. To investigate how the hysteretic transition is possibly linked to the density wave state, we examined the structural and electronic properties of the CDW at different temperatures along the loop, labeled as T1--T4 in Fig.~\ref{fig:nolockin}(a). No additional satellite peaks emerge in the $(HK0)$ diffraction plane as EuTe$_4$ is cooled from 300\,K to 30\,K [Fig.~\ref{fig:nolockin}(a), inset]. Furthermore, the CDW wavevector $q_\text{CDW}$ shows negligible variation within experimental uncertainty [Fig.~\ref{fig:nolockin}(c)]. The lack of temperature-dependent feature is echoed in the electronic dispersion, which displays no renormalization of the folded CDW band except for thermal broadening and a rigid shift along the energy axis [Figs.\,\ref{fig:nolockin}(b) and S3].

\begin{figure}[htb!]
	\includegraphics[width=\columnwidth]{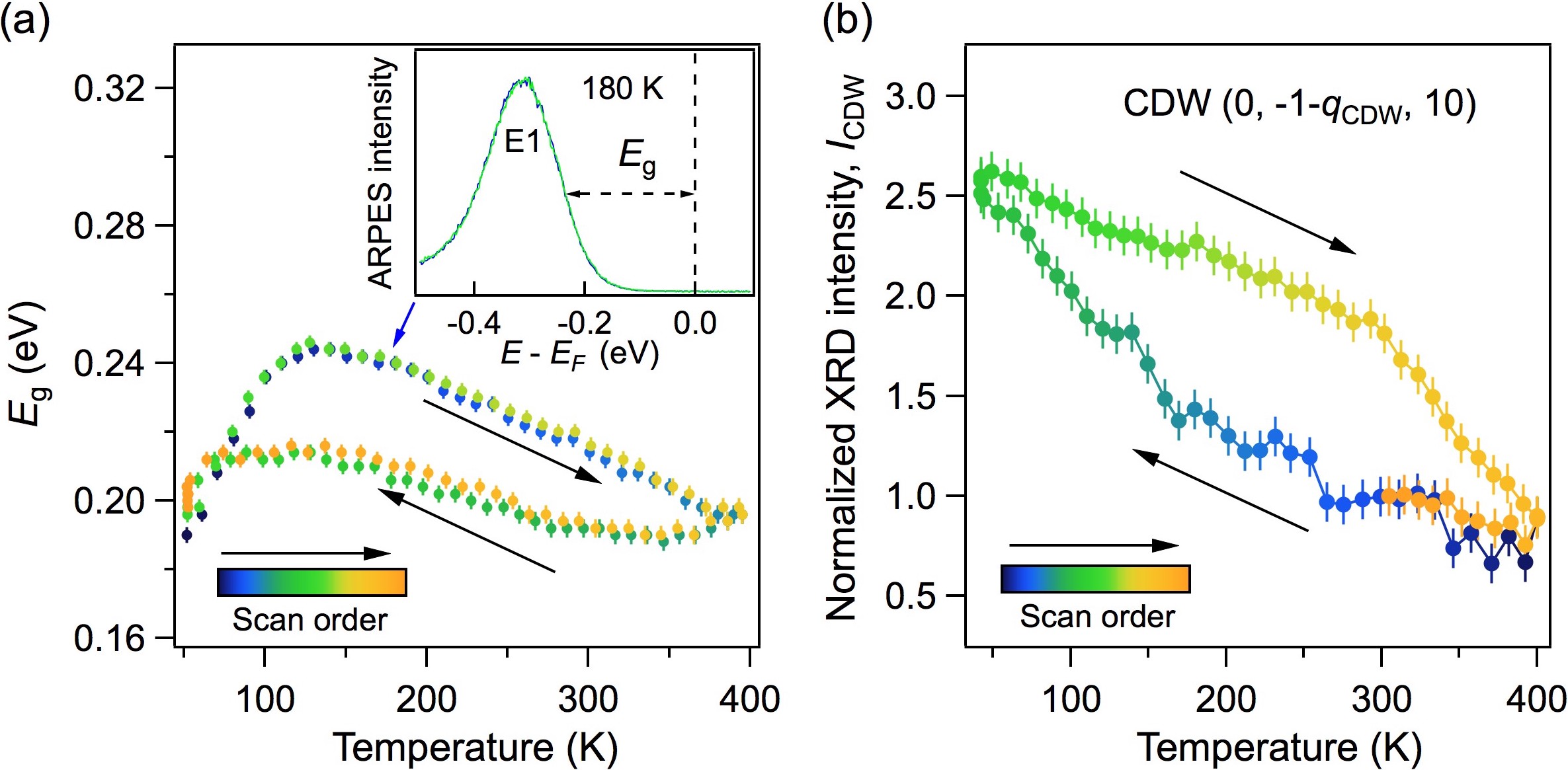}
	\caption{Giant thermal hysteresis within the CDW state. (a)~Leading edge gap $E_g$ from two thermal cycles 50\,K$\rightarrow$400\,K$\rightarrow$50\,K$\rightarrow$400\,K$\rightarrow$50\,K. $E_g$ is taken at its maximum value at momentum E1 [see Fig.~\ref{fig:intro}(f)]. The turn around 120\,K is interpreted as a change in the chemical potential. \emph{Inset}: two representative energy distribution curves at E1, measured at 180\,K in two successive heating branches. (b)~Temperature dependence of integrated intensity of a first-order CDW peak, normalized by its value at 300\,K at the end of a 400\,K$\rightarrow$42\,K$\rightarrow$400\,K$\rightarrow$300\,K sequence.}
\label{fig:hysteresis}
\end{figure}

Motivated by the energy shift of the band dispersion, we performed ARPES measurements between 50\,K and 400\,K to track the leading-edge gap $E_g$ in the high-symmetry $\bar{\Gamma}$--$\bar{\text{X}}$ direction. These experiments are difficult to perform because the sample surface can be easily contaminated by outgassing at elevated temperatures. Using a local heating technique \cite{Chen2019}, we observed no surface degradation, as evidenced by the excellent agreement between two thermal cycles (50\,K$\rightarrow$400\,K$\rightarrow$50\,K$\rightarrow$400\,K$\rightarrow$50\,K) performed in succession on the same sample [Fig.~\ref{fig:hysteresis}(a)]. Remarkably, the gap value $E_g$ exhibits hysteretic behavior with a similar temperature range as resistivity. An evolving leading-edge gap can be caused by changes in either the CDW gap size $\Delta$, or the chemical potential, or both. To distinguish these two contributions, we examine the temperature dependence of the first-order CDW peak, whose integrated intensity $I_\text{CDW}$ scales as $\Delta^2$ but does not depend on the chemical potential. As shown in Fig.~\ref{fig:hysteresis}(b), hysteresis is observed in $I_\text{CDW}$ but is negligible in the corresponding Bragg peak (Fig.~S4). Based on the values of $I_\text{CDW}$ and $E_g$ in the heating and cooling branches, we further estimate that the hysteresis in the leading-edge gap has a dominant contribution from the CDW gap instead of the chemical potential \cite{Remark1}. Taken together, our transport, photoemission, and diffraction experiments depict a unified phenomenology for the hysteresis loop: Compared to the cooling branch, the average CDW amplitude is stronger in the heating branch, leading to a larger energy gap and a higher resistivity value.

\begin{figure}[htb!]
	\includegraphics[width=\columnwidth]{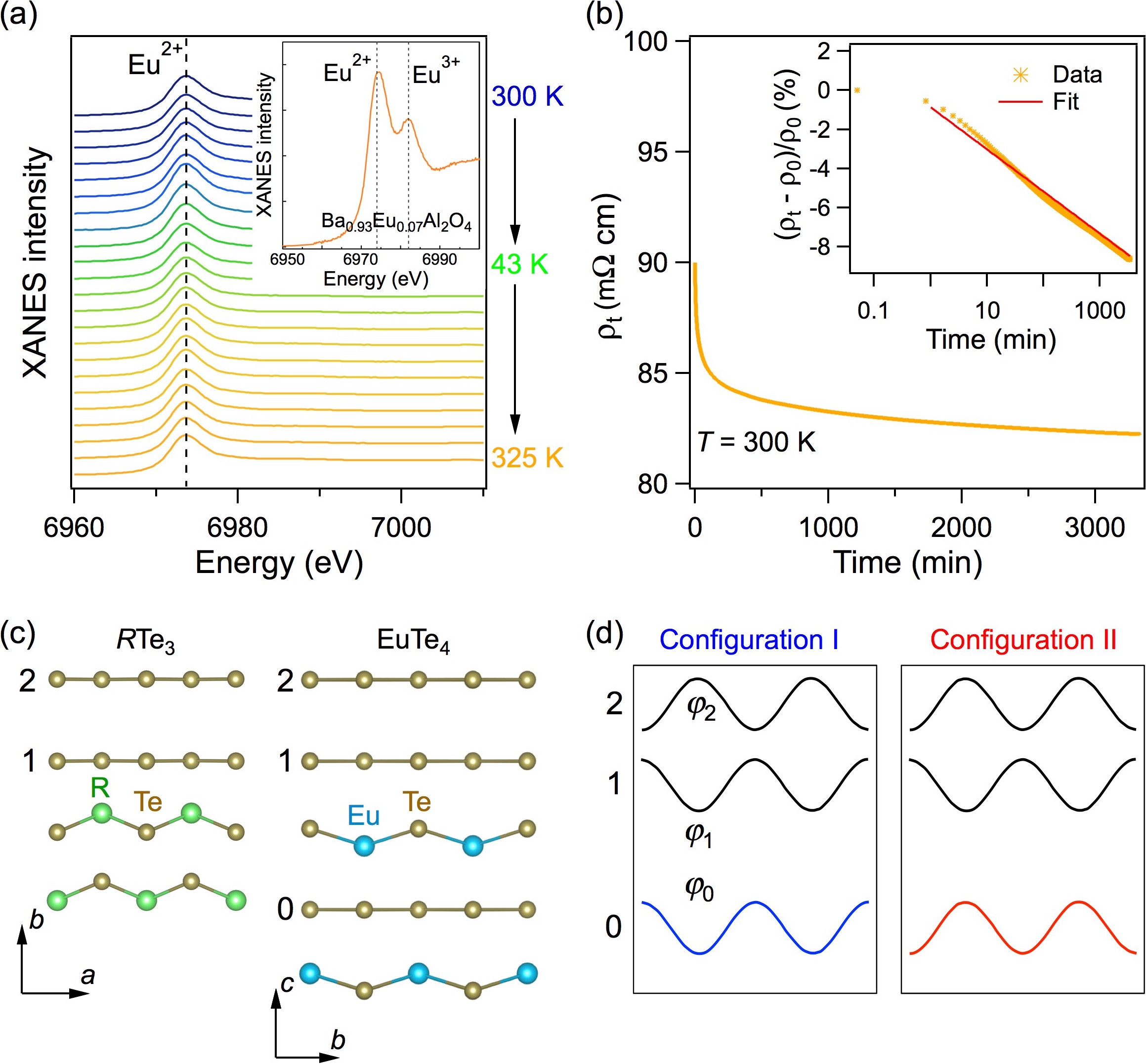}
	\caption{Candidate mechanisms of the hysteretic transition. (a)~Temperature dependence of the Eu $L_3$ x-ray absorption near edge structure (XANES) in EuTe$_4$. \emph{Inset}: XANES spectrum of Ba$_{0.93}$Eu$_{0.07}$Al$_2$O$_4$, adapted from ref.~\cite{Li2013}. (b)~Time dependence of electrical resistivity immediately after a thermal cycle of 400\,K$\rightarrow$20\,K$\rightarrow$300\,K [orange curve in Fig.~S6(c)]. \emph{Inset}: linear-log plot of the percentage change in resistivity relative to the first value of the time sequence ($\rho_0$). Red line is a logarithmic fit to $f(t) = A\log(t) + B$, where $A = -0.0215(2)$ and $B = -0.865(5)$ for a time window from 1~min to 3300~min. (c)~Schematic structures of $R$Te$_3$ and EuTe$_4$. Due to different space groups, the out-of-plane directions have different labels by convention. (d)~Illustrations of two candidate out-of-plane CDW configurations in EuTe$_4$. The relative CDW phase between Te monolayer and Te bilayer ($\phi\equiv\varphi_1-\varphi_0$) is assigned to 0 (I) or $\pi$ (II).}
\label{fig:summary}
\end{figure}

The presence of a hysteresis loop across more than 400\,K is surprising. The hysteresis occurs entirely within the CDW phase, which is to be contrasted to a hysteresis due to the metal-to-CDW transition. This is because the electronic structure does not show any remnant state inside the energy gap for all temperatures probed [Fig.~\ref{fig:nolockin}(b)]. Previous reports of thermal hysteresis in CDW compounds -- including quasi-1D Peierls-like systems \cite{Fleming1985}, quasi-2D transition metal dichalcogenides \cite{Wilson1975}, and 3D spinels \cite{Fleming1981} -- are ascribed to incommensurate-to-commensurate CDW transitions. In EuTe$_4$, however, the absence of detectable variation in its CDW wavevector renders this scenario improbable [Fig.~\ref{fig:nolockin}(c)]. For certain Eu-based compounds, another plausible mechanism for thermal hysteresis is the valence transition between the Eu$^{2+}$ and Eu$^{3+}$ states \cite{Martensson1982,Li2013,Rooymans1965}. To test this hypothesis, we performed x-ray absorption near edge spectroscopy on the Eu $L_3$ edge [Fig.~\ref{fig:summary}(a)]. All spectra collected from 43\,K to 325\,K feature a single peak at the Eu$^{2+}$ energy with no peak resolved from the background at the Eu$^{3+}$ energy. This observation is thus incongruent with a temperature-induced valence transition from Eu$^{2+}$ to Eu$^{3+}$. 

To gain more insights into the nature of the hysteresis, we turned to out-of-equilibrium response of the system. Specifically, we measured sample aging, a common characteristic of a hysteresis loop. Figure~\ref{fig:summary}b shows the temporal evolution of the in-plane resistivity at 300\,K, measured immediately after a 400\,K$\rightarrow$20\,K$\rightarrow$300\,K cycle. The resistivity follows a quasi-logarithmic decay over 3000\,min, suggestive of multiple metastable configurations in the system \cite{Mihaly1984,Duggan1985}. Within the measured time window, the resistivity drop is less than 10\%, in stark contrast to the order-of-magnitude difference between the cooling and warming branches. Therefore, regardless of microscopic details, the energy barriers between different metastable configurations are well in excess of $k_BT\sim26$\,meV at room temperature, leading to long-lived states. A tighter bound on the energy barrier $w$ can be obtained from the resistivity relaxation time, $\tau\gg3000$\,min, which implies $w \sim k_B T \ln ( \nu_o \tau ) \gtrsim 1$\,eV, where $\nu_o \sim 1$\,THz is the attempt frequency estimated from the typical phonon energy in this family of compounds \cite{Lavagnini2008,Hanggi1990}. An implication of this long relaxation is that any experimentally accessible cooling and warming rates are too fast in comparison with the characteristic timescale of a complete equilibration within the system. Indeed, we observed negligible difference between resistivity loops when we increased the temperature sweep rate from 0.15\,K/min to 10\,K/min [Fig.~S6(c)]. The lack of change between loops also contradicts to a scenario of a strong glass-like transition, which is expected to sensitively depend on the sweep rate and sample history \cite{Duggan1985,Berthier2011}.

Our combined equilibrium and nonequilibrium measurements point towards a rather unconventional hysteretic transition in EuTe$_4$, which is clearly distinguished from metal-to-CDW, incommensurate-to-commensurate, valence state, or strong glass-like transitions. An additional clue to understanding the giant thermal hysteresis in EuTe$_4$ is its conspicuous absence in the closely related $R$Te$_3$ family \cite{Dimasi1995,Gweon1998,Brouet2008,Malliakas2005,RuThesis,yaohong2006}. In both tetratellurides and tritellurides, CDW originates from the Te layers. The apparent difference between $R$Te$_3$ and EuTe$_4$ is that the former only has Te bilayers while the latter has additional Te monolayers [Fig.~\ref{fig:summary}(c)]. In EuTe$_4$, the bilayer and monolayer CDWs share the same wavevector but exhibit distinct Te distortions \cite{Wu2019}. 

The above comparison suggests that the hysteresis is related to the interaction between the bilayer and the monolayer CDWs. Specifically, their interaction is expected to most significantly modify the degree of freedom that remains degenerate for isolated bilayers or monolayers -- the relative phase between the planar CDWs. In this regard, we propose that the giant thermal hysteresis is due to the switching of the relative phase between the CDW orders in the two types of Te layers. As illustrated in Fig.~\ref{fig:summary}(d), density waves in adjacent monolayer and bilayer planes can either be in phase or out of phase. These two configurations result from the competition between the CDW coupling to the lattice and the Coulomb repulsion between the mono- and bilayer CDWs. On symmetry grounds \cite{Remark1}, the relative CDW phase $\phi$ enters the free energy as $F = F_0 + a(T)\cos\phi + b\cos^2\phi$, where $F_0$, $a$ and $b$ are phenomenological coefficients. For sufficiently small $a$, the free energy $F$ has two minima approximately shifted by $\pi$ (Fig.~S7). We propose the switching between these two minima to be the mechanism underlying the hysteresis. This mechanism entails two theoretically distinct scenarios discussed in detail in ref.~\cite{Remark1}. According to the first scenario, there could be a first-order phase transition due to a sign change of $a(T)$ at a particular temperature. The second scenario stipulates that the symmetry of the 3D CDW arrangement may impose $a=0$ in the entire temperature range, which leads to the Ising-like degeneracy of the two free energy minima and hence the hysteretic formation of corresponding CDW domains. The atypically large thermal width of the hysteresis would be particularly natural for this scenario. 

Adding to the preceding considerations, we highlight two microscopic factors that set apart the interplane CDW couplings in EuTe$_4$ from those in $R$Te$_3$. The divalence of the Eu ions implies that the Te planes neither accept nor donate electrons to the Eu-Te slab, so the Te mono- and bilayers remain nominally neutral. By contrast, Te planes in $R$Te$_3$ receive one electron per bilayer from the $R$-Te layer. In addition, the Coulomb coupling between CDWs in different Te layers is expected to be better screened by quasiparticles in $R$Te$_3$ than in EuTe$_4$ \cite{Staresinic2008}. This is because the Fermi surface in EuTe$_4$ is fully gapped by the CDW whereas there is a remnant Fermi surface in the CDW state of $R$Te$_3$. The charge neutrality of the Te layers and the reduced number of mobile carriers likely play an important role in setting the energy scale of the out-of-plane CDW coupling, leading to a delicate balance between different relative phases that underlies the giant hysteretic transition. 

Our multi-probe investigation of the incommensurate CDW in EuTe$_4$ demonstrates several peculiar features of the thermal hysteresis: (i)~it has a giant temperature span, (ii)~it is accompanied by large differences in the average CDW amplitude,
(iii)~it does not involve any change in the valence state or CDW commensuration within our experimental resolution, (iv)~no symmetry breaking is observed in the course of the transition, and (v)~it is rather insensitive to cooling or heating rate. These observations demonstrate an unconventional origin of the hysteresis. Our analysis suggests that the hysteresis may be a manifestation of the switching between different 3D configurations of the in-plane density waves. This switching, resulting from the delicate balance of interlayer couplings, points to a transition unique in quasi-2D systems, which is exclusively associated with out-of-plane orders as opposed to instabilities within each plane. Our findings not only expand the phenomenology of hysteretic transitions in broken-symmetry states, they also open the possibility of manipulating the rich internal structure of CDWs, raising numerous opportunities for device application that capitalizes on the wide temperature range of the metastable behavior.

\begin{acknowledgments}
We thank Patrick A. Lee, Liang Fu, Joseph G. Checkelsky, Ian R. Fisher, Linda Ye, Yang Zhang, Noah F. Q. Yuan, Jiarui Li, Anshul Kogar, Yu He, Bryan Fichera and Batyr Ilyas for fruitful discussions. We acknowledge support from the U.S. Department of Energy, Office of Science, Office of Basic Energy Sciences, DMSE (instrumentation and data taking), the National Science Foundation under Grant No. NSF DMR-1809815 (data analysis), and the Gordon and Betty Moore Foundation's EPiQS Initiative grant GBMF9459 (manuscript writing). A.Z. acknowledges support from the Miller Institute for Basic Research in Science. D.W. and N.L.W. acknowledge support from the National Natural Science Foundation of China (No.~11888101), and the National Key Research and Development Program of China (No.~2017YFA0302904). H.W. acknowledges support from the U.S. Department of Energy, Office of Science, Office of Basic Energy Sciences, Materials Sciences and Engineering Division, under contract No. DE-SC0012509. Work at the Advanced Photon Source and the Center for Nanoscale Materials was supported by the U.S. Department of Energy, Office of Science, under contract No.~DE-AC02-06CH11357, and the Canadian Light Source and its funding partners. This research was partly supported by the Army Research Office through Grant No.~W911NF1810316, and the Gordon and Betty Moore Foundation EPiQS Initiative through grant GBMF9643 to P.J.-H. (sample preparation and characterization). This work made use of the Materials Research Science and Engineering Center Shared Experimental Facilities supported by the National Science Foundation (NSF) (Grant No. DMR-0819762). This work was performed in part at the Harvard University Center for Nanoscale Systems (CNS), a member of the National Nanotechnology Coordinated Infrastructure Network (NNCI), which is supported by the National Science Foundation under NSF ECCS Award No. 1541959. Xijie Wang acknowledges support from the U.S. Department of Energy BES SUF Division Accelerator \& Detector R\&D program, the LCLS Facility, and SLAC under contract No.'s DE-AC02-05-CH11231 and DE-AC02-76SF00515 (MeV UED at SLAC). Research conducted at CHESS is supported by the NSF via awards DMR-1332208 and DMR-1829070. The work at SSRL is supported by the U.S. Department of Energy (DOE), Office of Science, Office of Basic Energy Sciences, Division of Materials Sciences and Engineering, under contract DE-AC02-76SF00515. D.H.L. was supported by the DOE, Office of Science, Office of Basic Energy Sciences, Division of Materials Sciences and Engineering, under grant DE-AC02-05CH11231. Y.B.H. acknowledges support from the National Key Research and Development Program of China (2017YFA0403401) and the National Natural Science Foundation of China (U1875192,U1832202). Y.W. acknowledges support from the National Science Foundation (NSF) award DMR-2038011.
\end{acknowledgments}

\end{document}